\documentclass[doublecol]{epl2} 
\usepackage{float}
\usepackage{subfloat}
\usepackage{subfig}
\usepackage{hyperref}

\title{ The need for statistical physics in Africa: \\ perspective and an illustration in   drug delivery problems }
\shorttitle{The need for statistical physics in Africa (perspective)} 

\author{Mtabazi G. Sahini\inst{1} \and Isaac Onoka\inst{1} \and Said A.H. Vuai\inst{1} \and \'Edgar Rold\'an\inst{2}   \and Hulda S Swai\inst{3}  \and \\ Daniel M. Shadrack\inst{3,4}\footnote{Corresponding author}}
\shortauthor{M. G. Sahini \etal}

\institute{                    
  \inst{1} Department of Chemistry, College of Natural and Mathematical Sciences, University of Dodoma,  338 Dodoma, Tanzania\\
  \inst{2} ICTP -- The Abdus Salam International Centre for Theoretical Physics, Strada Costiera 11, 34151 Trieste, Italy\\
  \inst{3} NM-AIST -- The Nelson Mandela African Institution of Science and Technology, P.O.Box 447 Arusha, Tanzania\\
   \inst{4} Department of Chemistry, Faculty of Natural and Applied Sciences, St John's University of Tanzania, 47 Dodoma, Tanzania
}

\abstract{
The development of statistical physics in Africa is in its nascent stages, yet its application holds immense promise for advancing emerging research trends on the continent. This perspective paper, a product of a two-week workshop on biophysics in  Morogoro (Tanzania), aims to illuminate the potential of statistical physics in regional scientific research. We employ in-silico atomistic molecular dynamics simulations to investigate the loading and delivery capabilities of lecithin nanolipids for niclosamide, a poorly water-soluble drug. Our simulations 
reveal that the loading capacity and interaction strength between lecithin nanolipids and niclosamide improve with increased lecithin concentrations. We perform a free-energy landscape analysis which uncovers two distinct metastable conformations of niclosamide within both the aqueous phase and the lecithin nanolipids. Over a simulation period of half a microsecond, lecithin nanolipids self-assemble into a spherical monolayer structure, providing detailed atomic-level insights into their interactions with niclosamide. These findings underscore the potential of lecithin nanolipids as efficient drug delivery systems. } 

\begin{document}

\maketitle

\section{Introduction}
The first ``settlements" of statistical physics in Africa date back from the end of the 20th century, that is around a century later than the first pioneering studies by  Boltzmann, Maxwell, etc. Over the last decades,  examples of key fundamental discoveries led by African statistical physicists  are becoming increasingly more frequent. For example, the so-called Mpemba effect, discovered in the 1960s by  Erasto Batholomeo Mpemba (Tanzania)~\cite{mpemba1969cool}, is inspiring cutting-edge  research by a wide range of  statistical physicists in the 2020s~\cite{klich2019mpemba,bechhoefer2021fresh,moroder2024thermodynamics,ibanez2024heating}.    
 Statistical physics is hence nowadays gaining momentum in Africa, with research expanding in  e.g.  nonlinear physics and applied fractional calculus in Cameroon and Egypt~\cite{ndzana2007modulational,khan2020modeling}, molecular dynamics (MD) simulations in Tanzania~\cite{shadrack2021luteolin}, and nonequilibrium statistical mechanics in South Africa~\cite{nyawo2017minimal}.  However,  most post-graduate training efforts---led mostly by the pan-African AIMS (African Institute for Mathematical Sciences)  institutes' network---did not catalyse yet statistical-physics topics within their rich curricula.  Over the recent past, several biophysics schools  training in statistical physics techniques have been organized in Kenya, Tanzania, and Malawi, in conjunction with a recent school at the EAIFR (East Africa Institute for Fundamental Research, Kigali, Rwanda), which fosters regional expertise. These efforts highlight the increasing need for statistical physics in solving real-world challenges. In parallel to these  efforts, young African scientists are expanding  online scientific networks in modern topics of statistical physics, such as the successful Machine Learning and Data Science  (MLDS) in Africa forum (\url{http://mldsafrica.co.za/}). 
 
\begin{figure*}
	\captionsetup{type=table}
	\centering
	\includegraphics[width=0.95\textwidth]{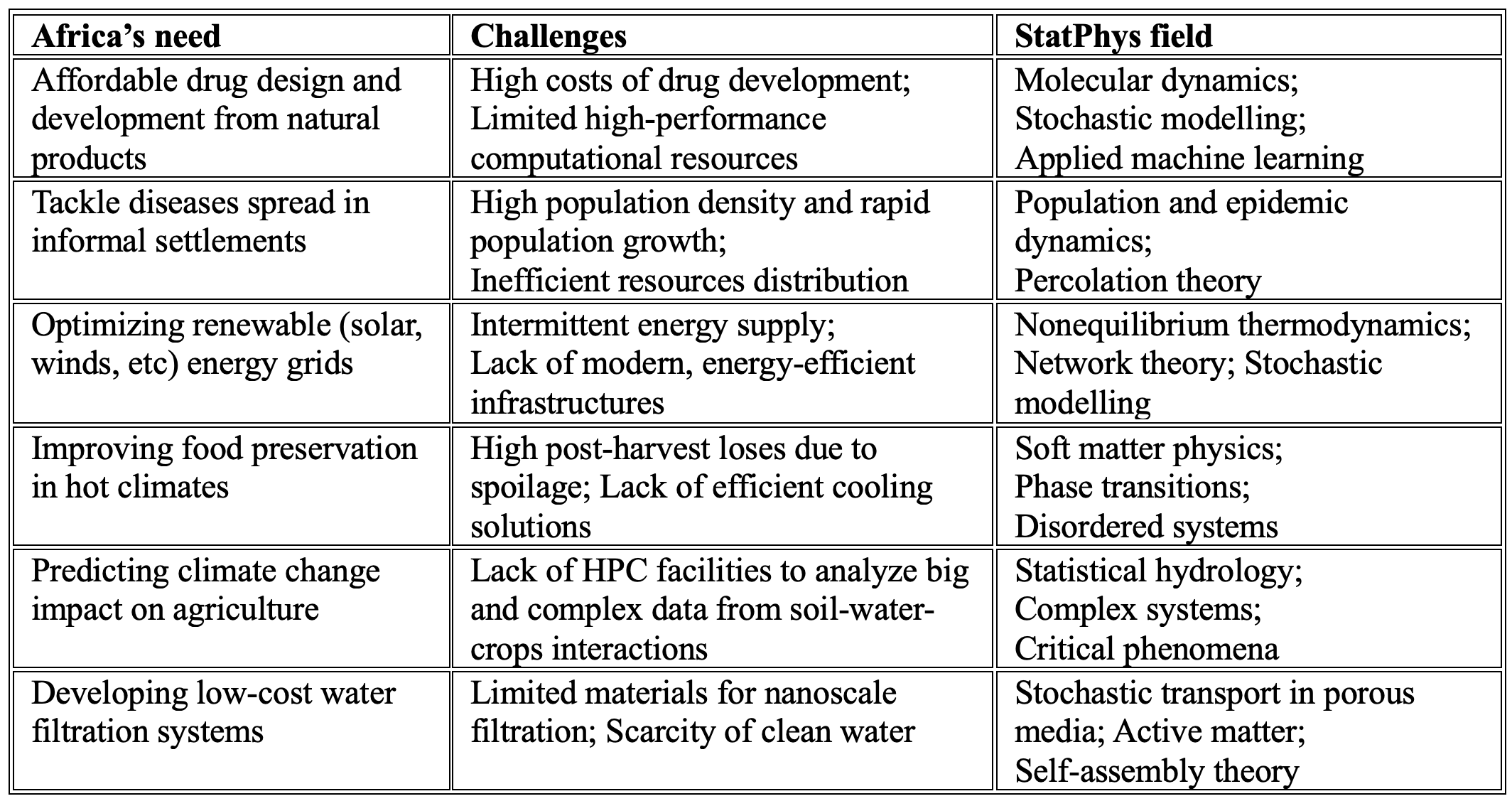} 
	\caption{Real-world potential future applications of statistical physics in Africa, highlighting key societal needs, associated challenges, and relevant knowledge areas within statistical physics  required to address them. }
	\label{fig001}	
\end{figure*}

 Beyond its role in advancing healthcare, statistical physics is crucial for addressing a wide range of challenges in Africa. As highlighted in Table~\ref{fig001}, various fields from drug discovery and renewable energy optimization to disease modeling and food preservation rely on statistical-physics-based approaches to overcome pressing challenges. For instance, molecular dynamics and machine learning are essential for affordable drug design, while non-equilibrium thermodynamics and network theory aid in optimizing renewable energy grids. Similarly, epidemic dynamics and percolation theory help model disease spread in informal settlements, whereas phase transition and soft matter physics play a key role in food preservation under extreme climates. Additionally, complex systems and stochastic processes are vital for predicting climate change impacts on agriculture and designing cost-effective water filtration systems. Despite these fields' ongoing growth in Africa, many of the areas outlined in the third column in Table~\ref{fig001}  require deeper expertise in statistical physics to develop sustainable solutions for the continent's needs. 

In this perspective paper, we  illustrate the critical role of statistical-physics tools in healthcare by employing molecular dynamics simulations to study drug delivery using lecithin nanoparticles. To this aim, we report results that are fruits of the training of  newcomers  in statistical physics in a two-week workshop in Morogoro (Tanzania) that  focused on how statistical physics can contribute to the study of phospholipid-based drug delivery systems, particularly lecithin nanolipids, which are gaining recognition as effective carriers for hydrophobic drugs. In this study, we explore the use of a computational tool from statistical physics to enhance the understanding of the role of lecithin nanolipids in drug delivery. 

Phospholipids are essential biomolecules known for their amphiphilic properties, which enable the formation of critical lipid bilayers to the structure of eukaryotic cell membranes. Comprising a hydrophilic phosphate head and two hydrophobic fatty acid tails, phospholipids self-assemble in aqueous environments driven by the repulsion between water and the hydrophobic tails. This versatile self-assembly leads to a plethora of phospholipid structures, including micelles and liposomes, which are highly valued in drug delivery systems for their ability to enhance drug bioavailability and facilitate solubilization, surfactancy, and controlled release \cite{singh2017phospholipids, hippalgaonkar2010injectable, rupp2010solubilization, allen2013liposomal,qaisrani2021phospholipids}.

\begin{figure}
	\centering
	\includegraphics[width=0.45\textwidth]{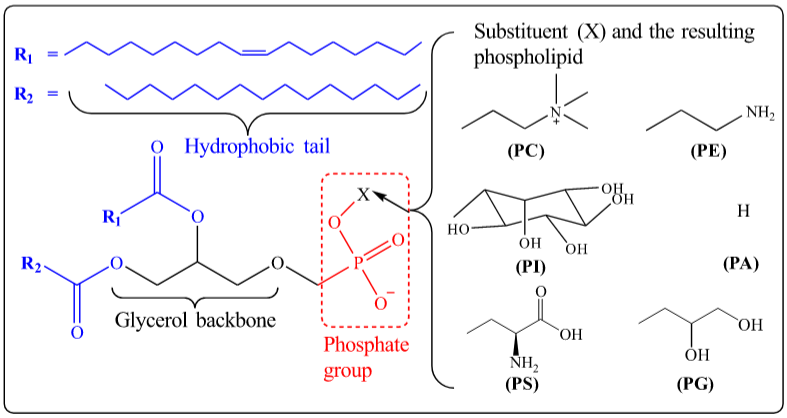} 
	\caption{ Illustration of the  chemical structure of various classes of glycerophospholipid lecithins. In the present study, the primary focus was on investigating the drug-loading capacity of lecithin nanoparticles utilizing phosphatidylcholine (PC). }
	\label{fig0}	
\end{figure}

Recent experimental advances have realized phospholipid-based drug delivery systems, such as micelles, liposomes, and solid-lipid nanoparticles, making substantial progress in clinical applications. Notable examples include Cleviprex \cite{hippalgaonkar2010injectable}, Valium \cite{rupp2010solubilization}, Doxil \cite{allen2013liposomal}, and Silybin phytosome \cite{bhattacharya2009phytosomes}, which have demonstrated successful results in treating various conditions. Phosphatidylcholine (PC) (Fig.~\ref{fig0}), the most prevalent phospholipid in mammalian cells, is particularly promising for drug delivery applications due to its amphipathic nature, which supports compatibility with both aqueous and lipid environments \cite{van2017critical, kidd2009bioavailability, constantinides2008advances}. The ability of PC to interact with lipophilic drugs, such as niclosamide a poorly water-soluble compound classified as a biopharmaceutical classification system (BCS) class II \cite{lin2016preclinical} makes it a valuable candidate for improving drug solubility and stability.

To address the challenge of low solubility of niclosamide, we employ molecular dynamics (MD) simulations, a computational method grounded in statistical physics, to investigate the interactions between lecithin (a phosphatidylcholine-based phospholipid) and niclosamide. Equilibrium statistical physics provides a framework for understanding the thermodynamic and kinetic behaviors of molecular systems at the atomic level. By applying this approach, we aim to gain insight into the molecular mechanisms underlying the loading and delivery of niclosamide by lecithin nanolipids.

The variation in lecithin concentrations impacts the loading capacity and stability of niclosamide within lecithin nanolipids, showing enhanced drug loading and more compact structures with increased lecithin. These findings underscore the effectiveness of using statistical physics in elucidating detailed interactions between lecithin and hydrophobic drugs, providing valuable atomic-level insights that complement experimental~research.

In this study, we highlight the potential of lecithin as a carrier for lipophilic drugs, by employing computational approaches to advance our understanding of drug-carrier interactions. By integrating statistical physics with molecular dynamics simulations, we offer a quantitative perspective on the molecular processes involved, paving the way for further experimental validation and optimization of phospholipid-based drug delivery systems.

\section{Computational methods}
\label{S:2}
\subsection{Molecular dynamics simulation}

In this study, we investigate the self-assembly process of lecithin nanolipid loaded with niclosamide through atomistic molecular simulations using Gromacs v2018 \cite{abraham2015gromacs} and the GROMOS 54a7 force field \cite{schmid2011definition}. The monomers of lecithin (CID:192817) and niclosamide (CID:4477) were obtained from the PubChem database \cite{kim2016pubchem} and prepared using the open babel tool \cite{o2011open}. Hydrogen atoms were added at pH 7.4, followed by energy minimization and conversion to the PDB file format. The coordinates and topology for lecithin lipid and niclosamide were generated using the PRODRG server~\cite{schuttelkopf2004prodrg}. It should be noted that although the PRODRG server has some topology quality limitations, we carefully examined and verified the topology and coordinates before starting the simulations.

Four different systems 
were prepared, each containing 0, 1, 4, and 128 lecithin (LEC) monomers interacting with a single niclosamide (NIC) molecule in independent simulation boxes:
\begin{itemize}
\item NIC 1:1 LEC, box size $(3\times 3\times 3)$ nm, 839 solvate molecules, production run $100$ns;
\item  NIC 1:4 NIC, box size $(3.05\times 3.05\times 3.05)$ nm, 742 solvate molecules, production run $100$ns; and
\item  NIC 1:128 LEC, box size $(10.13\times 10.52\times 10.30)$ nm, 31837 solvate molecules, production run $480$ns.
\end{itemize}
 These systems were solvated with the TIP3P water model \cite{gereben2011accurate}. Each system underwent energy minimization using the steepest descent algorithm, followed by two equilibration steps in the NVT and NPT ensembles, each lasting 500 ps, at 300 K and 1 bar. Temperature coupling was achieved using the v-rescale thermostat \cite{bussi2007canonical}, while pressure coupling was achieved using the Parrinello-Rahman barostat~\cite{parrinello1981polymorphic}. Subsequently, the equilibrated systems were subjected to production runs in the NPT ensemble at $300$K and $1$bar. During production, temperature and pressure were controlled using the velocity-rescale thermostat and the Parrinello-Rahman barostat, respectively.

We used the Particle Mesh Ewald (PME) method with a cutoff distance of 11\AA\, to handle long-range electrostatic interactions  and van der Waals interactions~\cite{darden1993particle, essmann1995smooth}. Covalent bonds were constrained using the P-LINCS algorithm \cite{hess2008p}, and a time step of $2$fs was used for integration in all systems. Periodic boundary conditions (PBC) were applied in all directions for all simulated systems.

\subsection{Free energy surface calculations}
The free-energy surface (FES) as the function of the selected coordinate was estimated as {$\Delta F = -k_BT \ln (P_x/P_0)$}, where, $k_B$ is the Boltzmann constant, $T$ is temperature and $P_x$ is the probability distribution (and $P_0$ a reference value) of the selected reaction coordinate $x$. Visualization and plotting was done using VMD and gnuplot tools.


\begin{figure}
	\centering

	\subfloat{\includegraphics[width=0.45\textwidth]{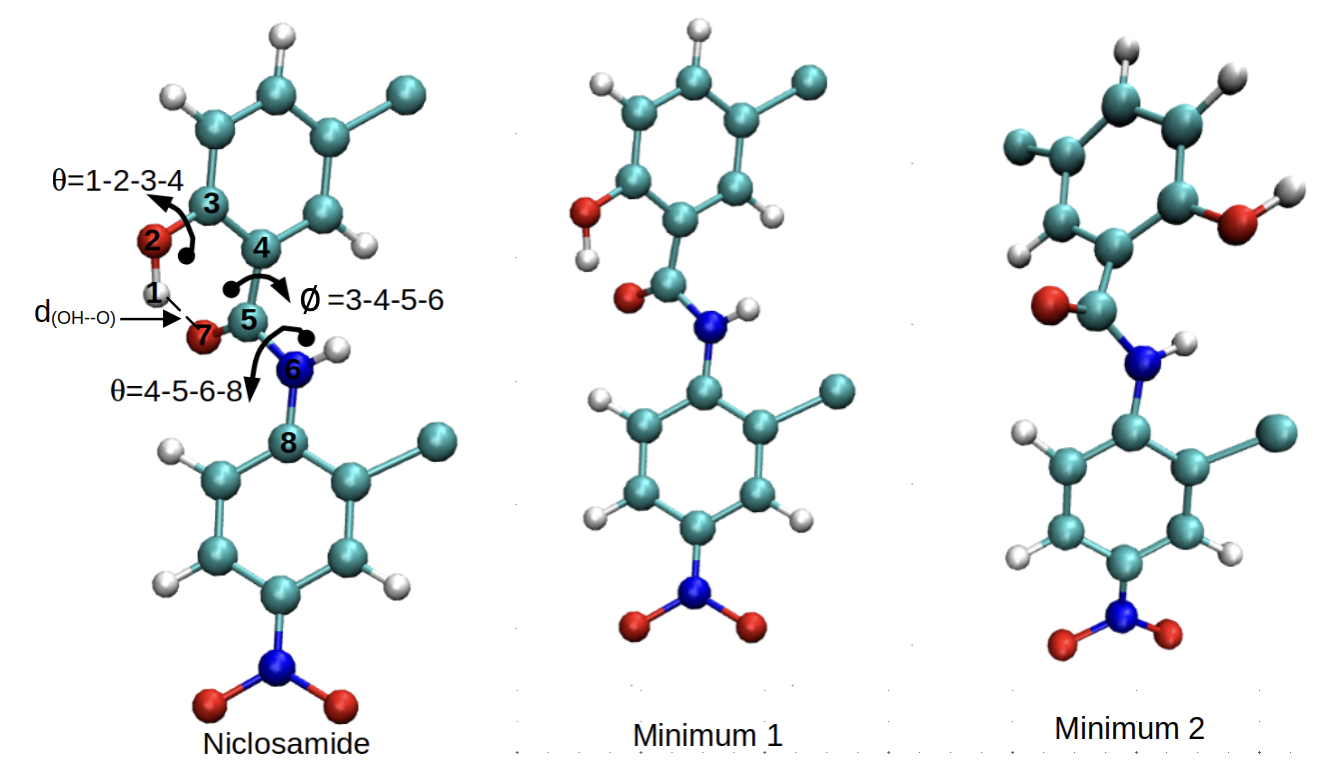}}
	\caption{Sketch of the chemical structure of niclosamide, together with  the keto-enol distance $d_{\rm OH-O}$  between atoms $1$ and $7$  and torsional angles $\theta$ and $\phi$. The central and right panel are sample  structures near the free-energy minimum 1 $(d_{\rm OH-O} \sim 2.8${\AA}) and at free-energy minimum 2 $(d_{\rm OH-O}\sim 4.5${\AA}), cf. Fig.~\ref{fig1}. Color code: red =  Oxygen, blue = Nitrogen, cyan =  Carbon, white = Hydrogen.}
	\label{fig2a}	
\end{figure}	

\section{Results and Discussion}
\subsection{Structural conformation of niclosamide in self-assembled lecithin nanolipids}

\begin{figure}
	\centering
	\subfloat[]{\includegraphics[width=0.48\textwidth]{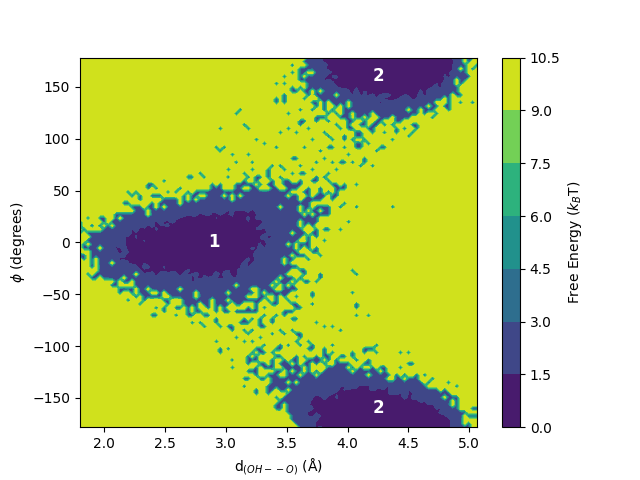}}\\
	\subfloat[]{\includegraphics[width=0.48\textwidth]{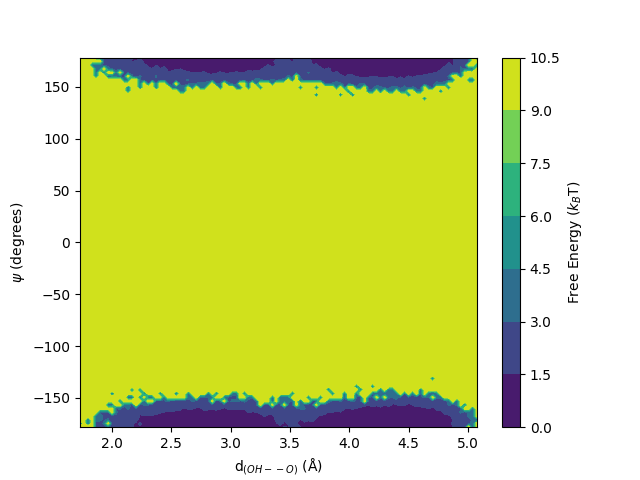}}
	
	\caption{2D free-energy surface of niclosamide (colormap) (a) $\Phi$ and $d_{\rm OH-O}$ indicating the conformation and orientation preference of keto-enol group. (b) $\Psi$ of the spacer region formed by atoms 4-5-6-8 and the distance $d_{\rm OH-O}$. In (a), 1 and 2 denote free-energy minima regions associated with the configurations sketched in Fig.~\ref{fig2a} for NIC 1:1 LEC concentration.}
	\label{fig1}	
\end{figure}

The conformational stability and orientational preference of niclosamide in water and nanolipids were explored by considering the keto-enol distance ($d_{\rm OH-O}$ 1,7) and torsional angles ($\phi$ 1-2-3-4), as reaction coordinates (Fig. \ref{fig2a}). The 2D free energy surface (FES) along the coordinates $(d_{\rm OH-O},\phi)$,  suggests that niclosamide prefers to have the enol group pointing away from the keto group and torsional angle $\phi$ switching between  $\phi=0$ and $\phi=$180$^{\circ}$ (Fig. \ref{fig1}a). The 2D FES thus suggests  two metastable conformational states of niclosamide in both water and nanolipids. To provide a further understanding of the origin of the distance fluctuations, we  also measured the  torsional angle $\psi$ at the spacer region formed by atoms 4-5-6-8 along with the keto-enol distance. We observed that niclosamide 
 fluctuates around  $\psi$ = 180$^{\circ}$ (Fig. \ref{fig1}b). The $\psi$ angles suggest that the deep minimum at $d_{\rm OH-O}$ $\sim$4.5~{\AA} is due to the orientation of the spacer region, which positions the OH group away from the keto group and results in intermolecular interactions between the enolic group and water/nanolipids. 

\begin{figure}
	\centering
	\subfloat[]{\includegraphics[width=0.45\textwidth]{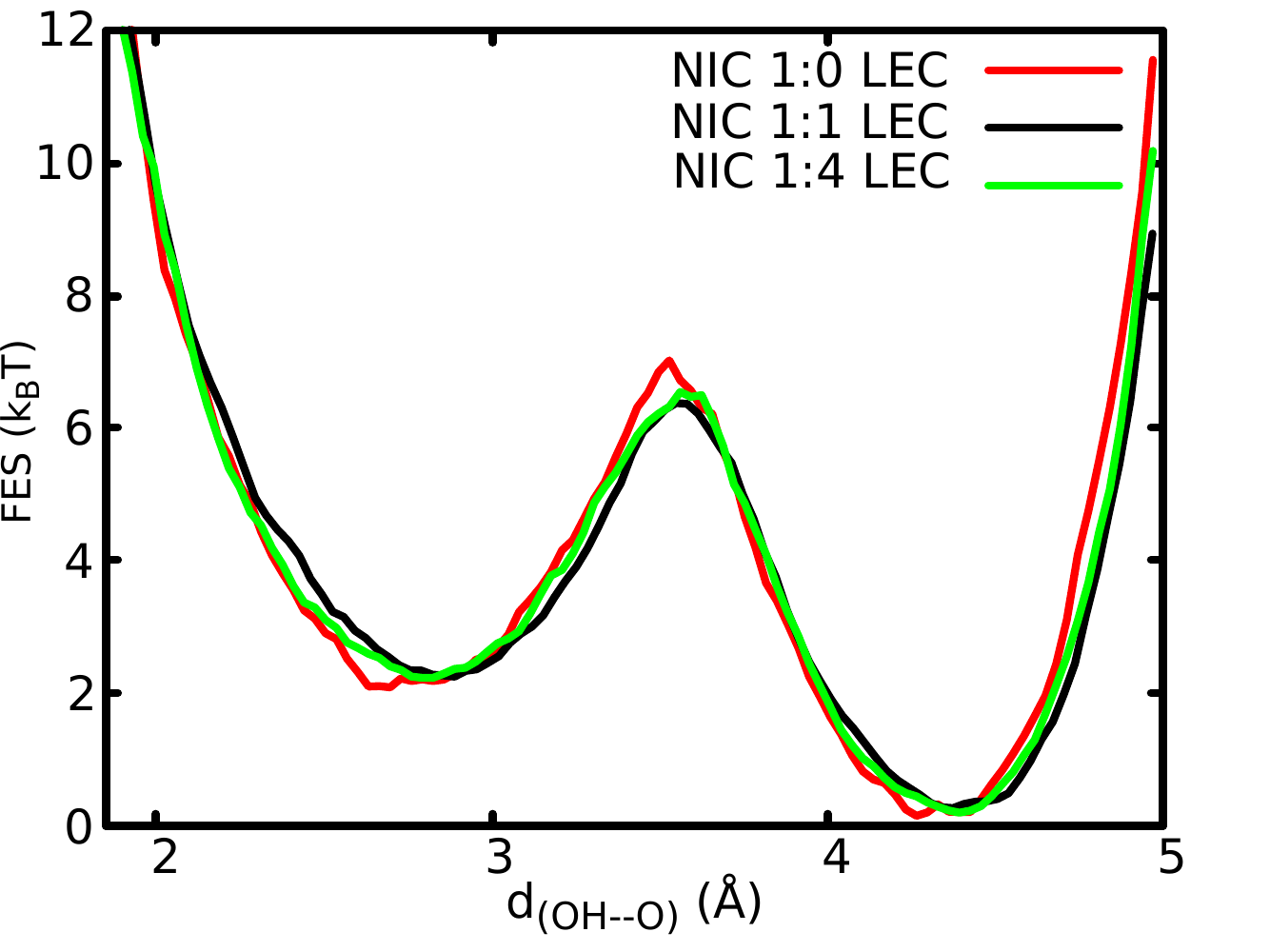}}\\
	\subfloat[]{\includegraphics[width=0.45\textwidth]{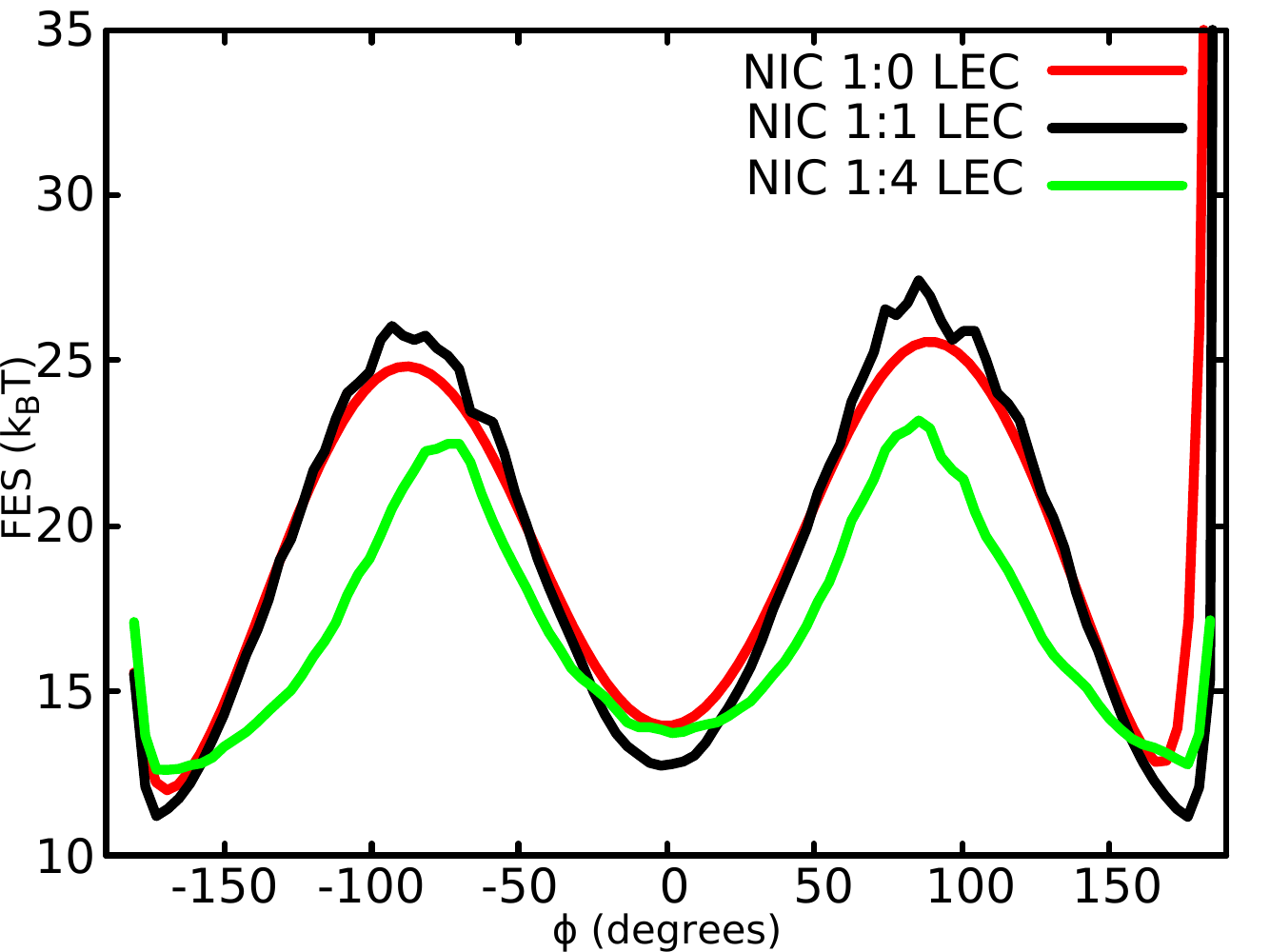}} 
	\caption{ 1D free-energy surfaces associated with the (a) keto-enol distances and (b) $\phi$ angle rotation for different niclosamide (NIC) vs lecithin (LEC) molecule ratios (see legend). The  chemical structure of niclosamide indicates the keto-enol distance ($d_{\rm OH-O}$ 1-7) along with structure at minimum ``1" ($d_{\rm OH-O} \sim 2.8 \mathrm{\AA}$) and minimum ``2" ($d_{\rm OH-O} \sim 4.5 \mathrm{\AA}$).}
	\label{fig01}	
\end{figure}

We also considered 1D free energy surfaces as the function of keto-enol distance ($d_{\rm OH-O}$, Fig.~\ref{fig01}a) and $\phi$ angle (Fig.~\ref{fig01}b). Figure~\ref{fig01} reveals an anisotropic free-energy landscape across the considered reaction coordinates: free-energy minima at~$d_{\rm OH-O}\sim 2.8\mathrm{\AA}$  and $d_{\rm OH-O}\sim 4.5\mathrm{\AA}$   with a kinetic barrier of $\sim 4k_{\rm B}T$ and minima at $\phi$ = 0$^{\circ}$ and $\pm180^{\circ}$ with kinetic barrier of  $\sim 10k_{\rm B}T$. Visual inspection of the free energy minimum 1 in Fig.~\ref{fig1}a shows the distance $d_{\rm OH-O} \sim 2.8 \mathrm{\AA}$ and $\phi$ = 0$^{\circ}$, corresponding to the $\alpha$ structure i.e enol hydrogen pointing the keto group. The free energy minimum 2 with $d_{\rm OH-O} \sim 4.5 \mathrm{\AA}$ and a rotation at $\phi$ = 180$^{\circ}$ corresponds to the $\beta$ structure, i.e enolic group pointing away from keto oxygen  (Fig. \ref{fig2a}).  On the other hand, the amide group prefers to form an intra-hydrogen bond with the enolic group, resulting in the formation and existence of the \textit{cis-trans} conformation (Fig. \ref{fig2a}). Our results are in good agreement with DFT calculations reported previously elsewhere for niclosamide in water, where it was shown that the $\alpha$ conformation is less stable than the counterpart\cite{romani2020properties}.

\subsection{Residence time and kinetics}
The residence time and kinetics are two key parameters commonly used in stochastic analysis that we use here  to describe niclosamide's interaction with  lipids. To understand how long it takes to niclosamide to change from one state to another i.e. from minimum 1 to 2, the residence time was estimated as per equation $\tau_{\rm 12} = \tau_{{\rm ex}} \exp( \Delta F/k
_{\rm B}T)$, where $\Delta F$ is the free energy difference between configuration A and transition state (TS), and $\tau_{{\rm ex}}$ is an average time taken from  minima 1 to 2 obtained over the simulation time, $\tau_{\rm 12}$ is the residence time taken to transition between a configuration  in the vicinity of minimum 1 to a configuration in the vicinity of minimum 2. See Table \ref{table1} for the estimated values of the mean residence time and transition rates obtained from a simulation in condition NIC 1:1 LEC. 
\begin{table}[H]
	\begin{center}
		\begin{tabular}{  |c |c| c| c | } 
			\hline
			Torsion angle &  $\tau_{\rm 12}$(ps) &k$_{{\rm off}}$(ps)  \\
			\hline
			1-2-3-4 & 1.5$\times$10$^1$ & 6.7$\times$10$^{-2}$ \\ 
			\hline
			4-5-6-8 & 7.4$\times$10$^4$ & 1.3$\times$10$^{-5}$ \\ 
			\hline
		\end{tabular}
		\caption{Mean residence time ($\tau_{\rm 12}$) and transition rate (k$_{{\rm off}}$) associated with niclosamide conformational changes  obtained from a simulation in condition NIC 1:1 LEC .\label{table1} }
	\end{center}
\end{table}

We observed that the residence time \( \tau_{\mathrm{ex}} \) associated with the rotation of  $\phi$ = {1-2-3-4} is of the order of approximately 3 ps. The residence times of two key torsion angles, $\phi$ = {1-2-3-4}  and  $\phi$ = {4-5-6-8}, which correspond to the observed conformations (minimum 1 and minimum 2, respectively, see also Fig.~\ref{fig2a}), were analyzed. As shown in Table~\ref{table1}, rotation around  $\phi$= {1-2-3-4}  is significantly faster than that around $\phi$ = {4-5-6-8}, with the latter  involving the spacer region. This reduced rotational rate is expected due to the presence of bulky substituents in the spacer region, which increase the free-energy barrier compared to the more flexible enol-containing segment. Indeed, as seen in the 1D free energy surface (FES) profiles in Fig.~\ref{fig01}, the conformational transition around  $\phi$ = {1-2-3-4} (minimum 1) is associated with a barrier of approximately \( \sim 4\,k_{\mathrm{B}}T \), while the barrier for rotation around $\phi$ = 4-5-6-8  (minimum 2) is much higher, approximately \( \sim 10\,k_{\mathrm{B}}T \). These differences in barrier heights are consistent with the observed disparity in residence times: the system transitions more readily from minimum 1 due to its lower barrier, while minimum 2 exhibits a slower exchange rate and longer residence time.

\subsection{Formation and structure stability of lecithin-niclosamide self-assembled nanolipid}
\begin{figure*}
	\centering
	\subfloat[]{\includegraphics[width=0.6\textwidth]{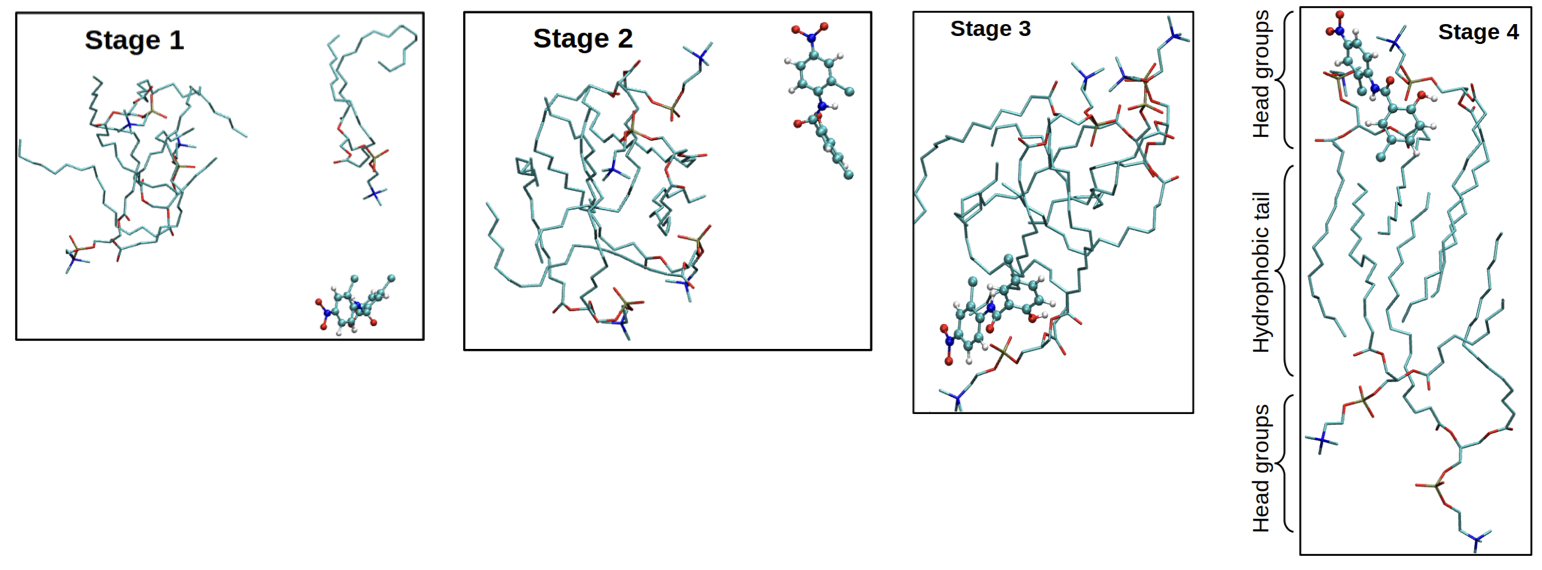}} \\
	\subfloat[]{\includegraphics[width=0.6\textwidth]{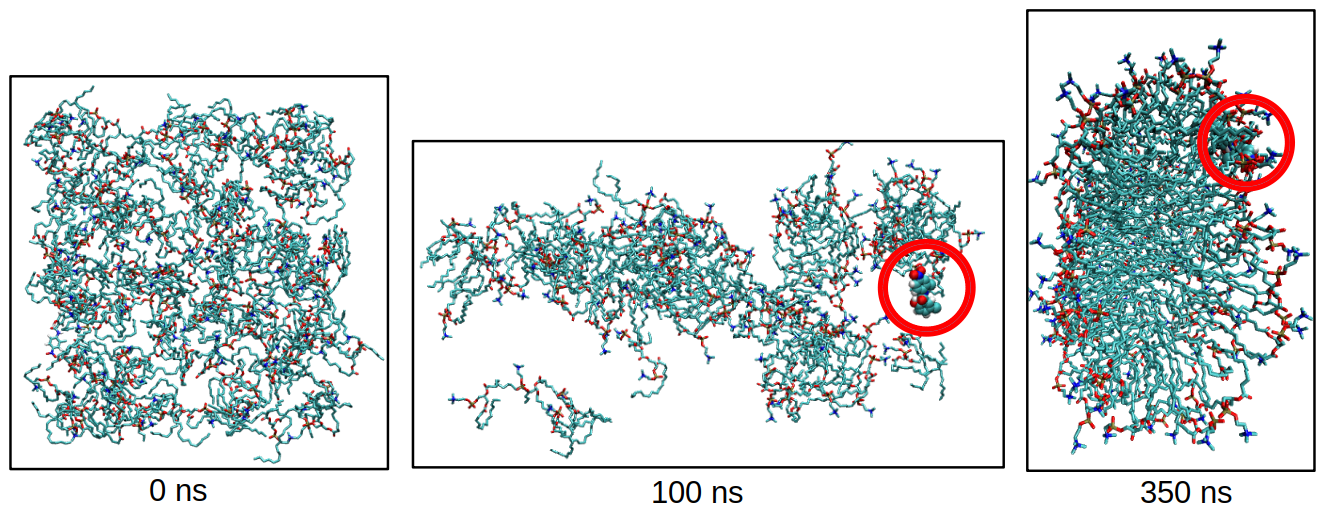}} \\
	\subfloat[]{\includegraphics[width=0.6\textwidth]{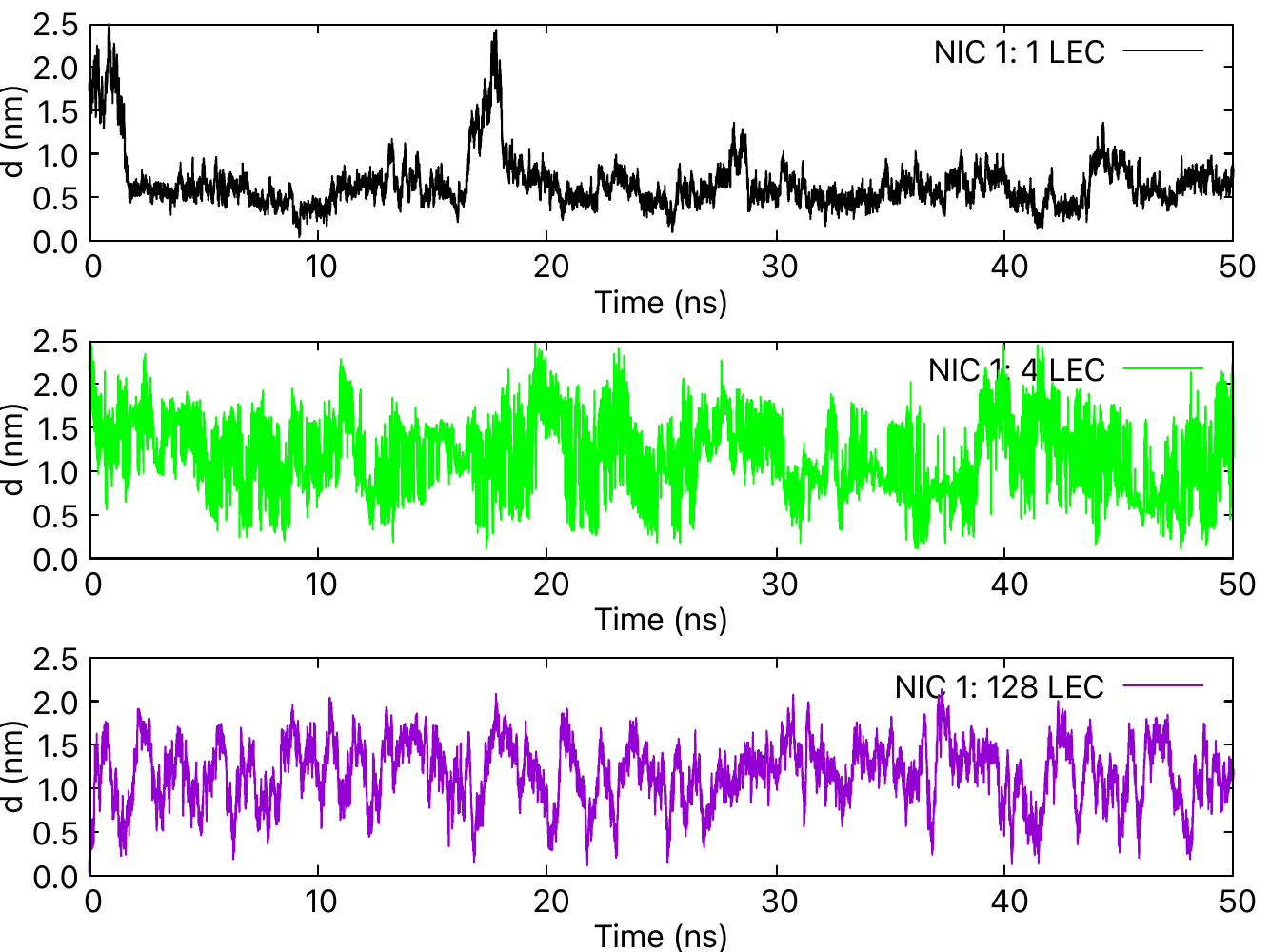}} 
	\caption{Formation of self-assembled lecithin monolayer with a single loaded niclosamide: (a) stages for self-assembly of 4 lecithin molecules. The self-assembly process maybe avalitatively described by 4 stages, niclosamide is loaded at stage 3 and the nanodisc formation has started, in stage 4 the lipids assembles into a monolayer/ bilayer.  (b) Snapshots of the elf-assembly process of 128 lecithin nanolipids, at 0 ns lecithin lipids are randomly placed in the simulation box, after 100 ns lipids starts to assemble themselves by forming small clusters with niclosamide loaded into one of the small cluster formed. During 350 ns a spherical monolayer nanolipid is formed with niclosamide loaded at the hydrophilic surfaces. The cross view of the spherical nanolipid is shown in the left panel.  The the red circle indicates the location of niclosamide in the nanolipid.  (c)  Centre of mass  distance between niclosamide and lecithin nanolipid as a function of time for  different lecithin concentrations. Water molecules are not shown for clarity. Atom colors are shown as C in green, O shown in red, P is shown in yellow, N is shown in blue.}
	\label{fig3f}	
\end{figure*}

Next, we investigated the self-assembly process of lecithin phospholipids and its ability to load a single niclosamide molecule in water  for different numbers  of lecithin phospholipids (1, 4 and 128). The self-assembly process of lecithin with 4 units  (Fig. \ref{fig3f}a-c), have a similar behavior with 128 units. In both cases, the self-assembly process started by joining together small monomers to form cluster of larger monomers, over time, the clusters sizes decreased and a formation of larger few clusters were observed. Such assembly process resembles soybean oil nanoemulsion system for curcumin delivery \cite{moghaddasi2018soybean}. During the self-assembly processes, we observed that some aggregates formed and quickly vanished; this process was repeated until a more stable thermodynamic lipid monolayer aggregate was obtained in hundreds of nanoseconds. The aggregation processes followed the same mechanism similar to other phospholipids reported by Hashemzadeh et al., \cite{hashemzadeh2020study}. Figure \ref{fig3f}b shows the self-assembly process starting from a randomly placed lecithin lipids in a simulation box with water. After tens to hundreds of nanoseconds the hydrophobic tail assembled them into micelle-like structure and then into lamellar-like. Finally, the head groups arranged into lamillar structure by forming a water-lipid interface with a spherical shape appearing similar to the scanning electron microscope (SEM) image for niclosamide-solid lipid nanoparticle (SLNs) reported in Figure 6 by Rehman et al., \cite{rehman2018fabrication}. 
Despite the lack of experimental data on niclosamide-lecithin nanolipid interactions and delivery, our findings are comparable to curcumin solid nanolipid systems \cite{moghaddasi2018soybean}. Although the current study did not go beyond investigating the permeation of nanolipid-loaded niclosamide through cell membranes, we hypothesize that the formed monolayer can act as a shuttle to carry and direct niclosamide to the membrane. This hypothesis is consistent with other poorly soluble molecules loaded in self-assembled lipid monolayers, such as danazol~\cite{kabedev2021molecular}.

\subsection{Stochastic motion of niclosamide in self-assembled lecithin nanospheres} 
During the simulation, the distance of niclosamide from lecithin lipids was measured (Fig. \ref{fig3f}c). Niclosamide slowly approached the surface of the lecithin lipid and eventually bound beneath the polar groups. During the first 10 ns of the simulation, niclosamide was observed to  stochastically bind and unbind; however, as the simulation progressed, niclosamide was loaded into the self-assembled lecithin monolayers, where it was observed to bind into the hydrophilic head groups (Fig. \ref{fig3f}b-c). The chloro-4-nitrophenyl ring always interacted with the head groups, while the hydroxybenzamide ring interacted with the hydrophobic tails.

We observe that the stochastic binding-unbinding dynamics of niclosamide to lecithin nanospheres was imprinted in the time series of the lecithin-to-niclosamide distance $d(t)$ (Fig. \ref{fig3f}c). Visual inspecting these time series, the reader may wander whether minimal coarse-grained stochastic models (e.g. Markov, Langevin) may be used to describe their fluctuations. Although inferring underlying stochastic models is a formidable task with fruitful applications~\cite{frishman2020learning,bruckner2021learning} we here focus on a simple discrimination task. In particular, we wander whether the time series  $d(t)$ may be described by a simple continuous-time two-level Markov jump process?  To do so, we evaluate numerical estimates of the autocorrelation function (ACF) associated with $d(t)$ for different lecithin concentrations (see symbols in Fig.~\ref{fig2str}).
\begin{figure}
	\centering
	{\includegraphics[width=3.4in]{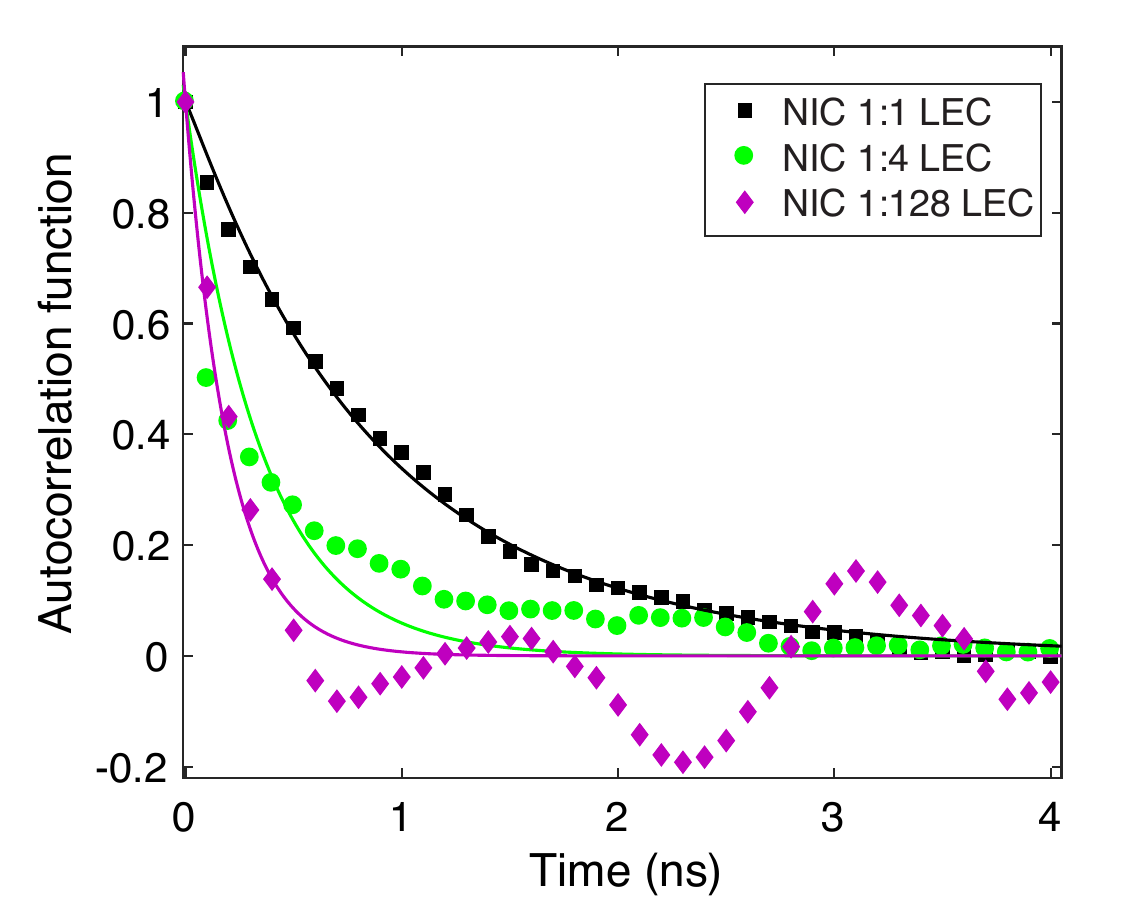}} 
	\caption{Autocorrelation function associated with the stochastic time series of the center-of-mass distance $d$ as between the niclosamide molecule and the lecithin nanospheres as a function of time (see Fig.~\ref{fig3f}c) for different lecithin concentrations  (symbols, see legend). The solid lines  are fits to exponential functions $\exp(-t/\tau)$. The fitting results and goodness of fit yield:  $\tau=(0.93\pm0.01)$ns $R^2=0.996$ (NIC 1:1 LEC),   $\tau=(0.35\pm0.01)$ns $R^2=0.7$ (NIC 1:4 LEC), and $\tau=(0.20\pm0.01)$ns $R^2=0.8$ (NIC 1:128 LEC). }
	\label{fig2str}	
\end{figure}
Our numerical estimates  reveal that, for low lecithin concentrations  (NIC 1:1 LEC), the ACF can be well approximated by an exponential decay $\exp(-t/\tau)$ with correlation time $\tau\sim 1$ns. On the other hand, increasing the lecithin concentration, the ACF displays damped oscillations, and thus cannot be accurately described by a simple exponential decay (see caption in Fig.~\ref{fig2str}). Moreover, such oscillations result of attachment/detachment events become quicker the larger is the concentration of lecithin. This leads us to conclude that the dynamics of the center of mass distance cannot be simply described by a Markovian two-level system at large lecithin concentrations, but rather non-trivial memory and inertial effects should be considered for accurate modeling.   

\begin{figure}
	\centering
	\subfloat[]{\includegraphics[width=3.0in]{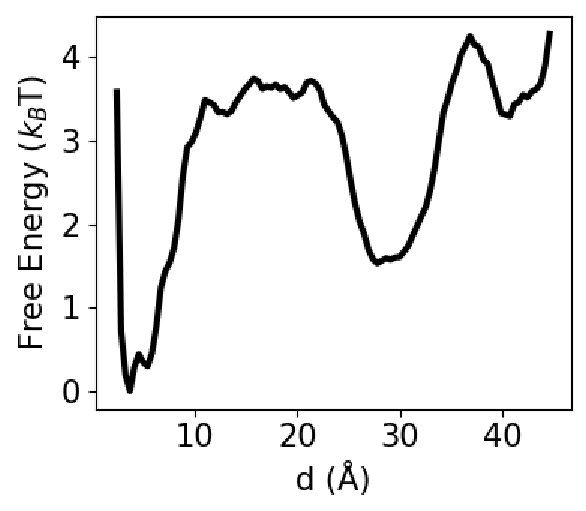}}\\
	\subfloat[]{\includegraphics[width=3.25in]{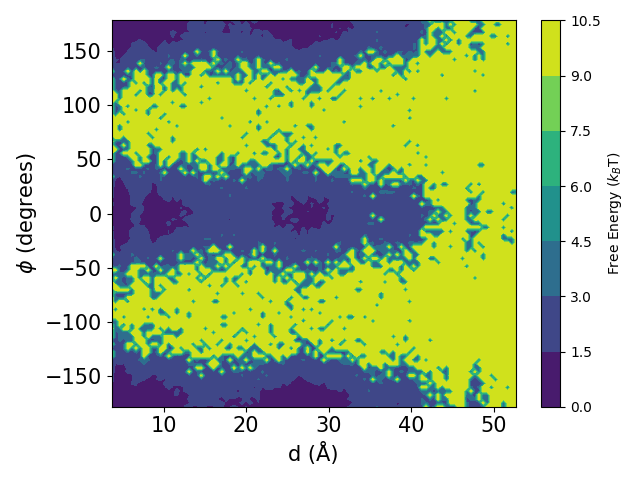}}
	\caption{(a) 1D free energy surface for NIC 1:128 LEC self-assembly process,  (b) 2D free energy surface as the function of distance and torsion angle. The free energy surface for the conditions NIC 1:1 LEC and NIC 1:4 LEC are not shown as they exhibited an analogous structure with two local minima separated by a kinetic barrier.}
	\label{fig2}	
\end{figure}

To investigate deeper the non-trivial statistics of the center-of-mass distance $d$ between lecithin and niclosamide we report in Fig.~\ref{fig2} a numerical analysis of  at 1D (Fig.~\ref{fig2}a) and 2D FES (Fig.~\ref{fig2}b)  at large lecithin concentratino (NIC 1: 128 LEC). For the 2D FES, we focused on its dependency  as a function of the distance $d$  and the angle of torsion $\phi$ = 1-2-3-4, which provides details on the binding and unbinding (release) process of niclosamide from lecithin lipids. The minimum free energy at a distance of $\sim$ 5 {\AA} denotes a bound state of niclosamide in lecithin phospholipids (Fig. \ref{fig2}a). The 2D free energy further provides insight into the (un)binding orientation of niclosamide in the lipid monolayer. Figure \ref{fig2}b shows that when the niclosmide in its bound and unbound state exhibited a rotation at 0 and $\pm$180$^{\circ}$, which suggests the existence of the \textit{ cis-trans} conformation, however, the \textit{trans} conformation is mostly dominant.

\subsection{Structural size of lecithin-niclosamide nanolipid}
 Readers may now wander what is the typical size of lecithin nanospheres during the assembly process? To answer this question, we assessed 
the structural size of the nanolipid complex by measuring the radius of gyration (Rg) over the simulation time. Rg is a key parameter that provides insights into the size and compactness of nanolipid-drug complexes. It has been widely used to characterize nanolipid structures \cite{fernandez2017computer} and assess lipid membrane compactness \cite{hashemzadeh2020study}. In this study, the Rg was calculated as  Rg(t) = $(\sum_{i}m_i \vert\vert \textbf{r}_i(t) \vert\vert ^2/ \sum_{i}m_i)^{1/2}$, where $m_i$ and $\textbf{r}_i(t)$ is the mass and position of each \textit{i}$-$th atom, respectively, with respect to the center of mass of the molecule.  The free energy surface (FES) as a function of Rg was obtained for lecithin-niclosamide nanolipid systems (Fig. \ref{fig2str}). A notable increase in Rg was observed; for instance, the system containing four lecithin monomers exhibited a two-fold size increase compared to a single monomer. The observed Rg increase was proportional to the number of monomers, aligning with findings from Fernandez et al. \cite{fernandez2017computer}, where the system size expanded with an increasing number of particles.
Furthermore, as system size increased, self-assembly and the formation of spherical-like structures required more time. This delay occurred because monomers repeatedly associated and dissociated from the growing cluster until reaching a more thermodynamically stable state. As shown before in Fig.~\ref{fig3f}a, self-assembly is a multistage process. However, once the lecithin nanolipids self-assembled, they became more compact, as indicated by the lower minimum free energy at 0.5 nm and 1 nm for systems with one and four lecithin nanolipids, respectively.

\begin{figure}
	\centering
	{\includegraphics[width=3.4in]{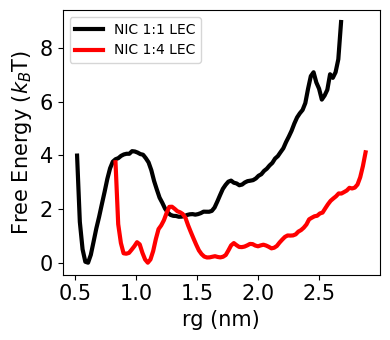}} 
	\caption{1D Free energy as a function of the radius of gyration (rg)  of the lecithin nanosphere interacting with a single niclosamide molecule. The two lines correspond to  1 (NIC 1:1 LEC) and 4 (NIC 1:4 LEC) lecithin nanolipid units in the simulation (see legend), revealing that the mean radius of gyration increased with the lecithin concentration in the aqueous environment.  The data associated with  128 lecithin molecules (NIC 1:128 LEC) displayed a similar trend and is not shown for the sake of visualization.}
	\label{fig2str}	
\end{figure}

\section{Conclusion}
\label{S:4}
The early stages of statistical physics development in Africa starting in the end of the 20th century have witnessed a major interest for applied science in lieu of fundamental aspects as in the foundations of the field in the 19th century.  In this perspective article, we reported the output of a postgraduate training activity in Tanzania, focusing on a soft matter biophysics application of computational tools brought from statistical mechanics. The results reported here suggest that lecithin nanoparticles exhibit significant promise as carriers for hydrophobic and poorly water-soluble drugs, with niclosamide serving as a key example in this study. Through molecular dynamics simulations, we have illustrated with molecular-dynamics simulation lecithin's potential to enhance drug solubility and stability. These simulations, grounded in statistical physics principles, offer critical insights into the self-assembly process of lecithin nanolipids, revealing their formation into stable monolayer structures that can efficiently load and deliver niclosamide. The application of statistical physics allows us to understand the thermodynamic and kinetic aspects of these interactions at an atomic level, providing a robust framework for evaluating the behavior of drug carriers. Despite these advancements, empirical validation through in vitro and in vivo experiments remains essential to confirm the practical stability and efficacy of lecithin nanolipids.

 This study not only highlights the potential of lecithin as a drug delivery system but also underscores the importance of integrating statistical physics with experimental research to fully realize and optimize the capabilities of nanolipid carriers in drug delivery applications which belong to some of the  pressing needs and challenges of the African scientific community as highlighted in Table~\ref{fig001}. Notably, such pressing challenges seem yet far away from current mainstream of statistical physics in the developed world (e.g. high-performance machine learning, quantum computing) which often require expensive and energetically-costly resources.  We expect our work to motivate new generations of 
 scientists to further develop the power of statistical physics in life science applications  through innovative pan-African initiatives. 

\acknowledgments
All authors acknowledge the financial support by the University of Dodoma through SAS funding program.  E. R. acknowledges
 financial support from PNRR MUR project
No. PE0000023-NQSTI. The authors declare the usage of solely purely-human cognitive processes in writing all the manuscript, with the exception of the composition Table I which was aided by  the ``oracular power" of DeepSeek AI (artificial intelligence) tool.

\bibliographystyle{ieeetr}       



%
%
%
%
%

\end{document}